\documentclass{INTERSPEECH2023}

\interspeechcameraready

\usepackage{hyperref}
\usepackage[hyphenbreaks]{breakurl}

\usepackage{enumitem}
\setlist[itemize]{align=parleft,left=0pt..1em}

\usepackage{listings}
\usepackage{xcolor}

\usepackage{amsmath,graphicx}
\usepackage{color}
\usepackage{setspace}
\hypersetup{
   breaklinks=true,   
   pdfusetitle=true,  
}

\usepackage{color}
\definecolor{codegreen}{rgb}{0,0.6,0}
\definecolor{codegray}{rgb}{0.5,0.5,0.5}
\definecolor{codepurple}{rgb}{0.58,0,0.82}
\definecolor{backcolour}{rgb}{1.00,1.00,1.00}
\lstdefinestyle{mystyle}{
    backgroundcolor=\color{backcolour},   
    commentstyle=\color{codegreen},
    keywordstyle=\color{magenta},
    numberstyle=\tiny\color{codegray},
    stringstyle=\color{codepurple},
    basicstyle=\ttfamily\footnotesize,
    breakatwhitespace=false,         
    breaklines=true,                 
    captionpos=b,                    
    keepspaces=true,                 
    numbers=left,                    
    numbersep=5pt,                  
    showspaces=false,                
    showstringspaces=false,
    showtabs=false,                  
    tabsize=2
}
\title{Real time spectrogram inversion on mobile phone}
\name{Oleg Rybakov, Marco Tagliasacchi, Yunpeng Li, Liyang Jiang, 
Xia Zhang, Fadi Biadsy}

\address{Google Research}
\email{\{rybakov,mtagliasacchi,yunpeng,jiangliyang,xiaz,biadsy\}@google.com}

\usepackage{setspace}

\begin{document}


\maketitle
 
\begin{abstract}
We present two methods of real time magnitude spectrogram inversion: streaming Griffin Lim(GL) and streaming MelGAN. We demonstrate the impact of looking ahead on perceptual quality of MelGAN. As little as one hop size (12.5ms) of lookahead is able to significantly improve perceptual quality in comparison to its causal version. We compare streaming GL with the streaming  MelGAN and show different trade-offs in terms of perceptual quality, on-device latency, algorithmic delay, memory footprint and noise sensitivity. For fair quality assessment of the GL approach, we use input log magnitude spectrogram without mel transformation. We evaluate presented real time spectrogram inversion approaches on clean, noisy and atypical speech. We specified conditions when streaming GL has comparable quality with MelGAN: noisy audio and no mel transformation. Streaming GL is 2.4x faster than real time on the ARM CPU of a Pixel4 and it uses 4.5x times less memory than MelGAN.
\end{abstract}

\section{Introduction}
\label{sec:intro}

Spectrogram inversion is an important task for Speech-to-Speech~\cite{Parrotron, SequenceToSequence, Translation} and Text-to-Speech~\cite{WangSSWWJYXCBLA17, NaturalTTS, DeepVoice3} models, where it is often referred to as waveform generation (or vocoder). This is often the last step of the audio processing pipeline that converts a magnitude spectrogram to audio samples. Griffin-Lim~\cite{GriffinLim} is the standard signal processing based approach for solving this task. It is an iterative method that requires processing of the whole audio sequence, so it can not run in streaming mode. Several methods were proposed to run GL in streaming mode: RTISI~\cite{RTISI}, RTISI-LA~\cite{RTISI-LA}. A combination of GL with deep neural network~(DNN) was proposed in ~\cite{DeGL} to balance the quality of the reconstructed signal with the computational load depending on the target application.

Neural network based vocoders have become a popular solution because they can generate speech at a higher quality than GL. A notable example is WaveNet~\cite{WaveNet}, which generates high quality speech, but due to its autoregressive nature has limitations on real time applications. Several improvements were proposed in FFTNET~\cite{FFTNET}, WaveRNN~\cite{WaveRNN}, LPCNet~\cite{LPCNet}. Non-autoregressive models were proposed in ParallelWaveNet~\cite{ParallelWaveNet} and WaveGlow~\cite{WaveGlow}. The latter  is characterized by a large model size, which can limit its deployment on mobile phones. Another class of non-autoregressive models is based on GANs~\cite{GAN}: \cite{MelGAN, HIFI, PARALLELWAVEGAN, AdversarialGeneration, GenerativeAdversarial}.

Previous works mostly focused on mel-spectrogram inversion. Instead, in this paper we focus on linear spectrogram inversion, i.e., without mel transformation, as described in Section~\ref{experiments_subjective}. There are two reasons for that: GL can generate high quality audio and linear spectrograms are used in real production speech-to-speech models, such as for example Parrotron~\cite{Parrotron}.

In this work, we consider on-device audio signal reconstruction from magnitude spectrograms. It has to be speaker independent, have minimal memory footprint, run faster than real time and have latency control at run-time. All above are the reasons we propose streaming GL (based on RTISI-LA~\cite{RTISI-LA}). Even though it uses standard signal processing approach we show that it has comparable quality with a model based on GAN\cite{MelGAN, HIFI} on noisy audio and real production Parrotron~\cite{Parrotron} application. We selected MelGAN as a baseline for streaming vocoder design because it is one of the state of the art and it is fully convolutional, so that we can execute it on multiple input samples in parallel even in streaming mode (for example LPCNet~\cite{LPCNet} or WaveRNN~\cite{WaveRNN} are also streamable but they have to be executed sequentially as streaming GL). In~\cite{MelGAN} it was already demonstrated that non streaming MelGAN can run faster than real time on CPU/GPU and outperforms non streaming GL, but there are no papers comparing streaming GL with streaming neural vocoder (e.g. with streaming MelGAN) and there is no subjective evaluation of streaming GL (e.g.~\cite{RTISI-LA} reported only SNR evaluation). Our paper fills these gaps.

\noindent \textbf{Main contributions:}

1 We propose our redesign of RTISI-LA~\cite{RTISI-LA} in TensorFlow and call it streaming GL (open sourced in Figure~\ref{fig:GL}). It allows us to benchmark streaming GL on any kind of hardware (CPU, TPU or GPU). For example on ARM CPU of Pixel4 we show several performance advantages of streaming GL over streaming MelGAN: it is more than 2.4x faster than real time and its memory usage is 4.5x times smaller than streaming MelGAN.

2 We present a streaming MelGAN and explore the perceptual quality of causal and non causal (with lookahead) versions. We show that a streaming MelGAN with only one hop delay (12.5ms)  lookahead, outperforms GL approaches and a strictly causal MelGAN on clean audio. We compare it with non streaming base model~\cite{MelGAN} (it has 187ms delay) and show trade-offs between subjective quality and delay.

3 We present subjective and objective evaluation of streaming GL and show that it has comparable quality with streaming MelGAN on noisy audio and production Parrotron application.

\section{Model architectures} \label{models}

All the models considered in this paper receive as input a log magnitude spectrogram generated according to the following steps:

1 The raw audio samples at 16kHz sampling frequency are processed by means of a pre-emphasis filter~\cite{HTK} with the coefficient set to 0.97;

2 The STFT is computed with FFT size \textit{fft\_size}=2048, frame size equal to 50ms (\textit{frame\_size}=800 samples@16kHz), frame step (a.k.a. hop size) equal to 12.5ms (\textit{frame\_step}=200 samples@16kHz) and Hann windowing~\cite{Hann};

3 The complex-valued STFT is converted to a real-valued spectrogram by computing the magnitude of each STFT coefficient;

4 The magnitude spectrogram is processed with a logarithmic compression function which is applied element-wise with added \textit{delta}=1e-2.





The resulting log-magnitude spectrogram is fed as input to a vocoder, which processes it in streaming mode and generates output audio frame in the time domain with length equal to 12.5ms (200 samples).

\subsection{Streaming Griffin-Lim algorithm}\label{subsectionGL}
We hypothesize that streaming GL can have performance advantages (latency and  model size) over streaming MelGAN. We also would like to compare streaming GL with streaming MelGAN on subjective and objective quality metrics (there is no such previous research).

The original design of causal streaming GL (also RTISI~\cite{RTISI}) and its non causal streaming version with look ahead RTISI-LA~\cite{RTISI-LA} can requires design of a special kernel (for example written in c++). This kind of kernel is hard to benchmark on different hardware vs a neural vocoder model which is written in TensorFlow or PyTorch. For side by side comparison it is important that competing approaches are compared in the same environment with the same deep learning framework. That is why in this section we present our design of ~\cite{RTISI-LA} in TensorFlow.

Our streaming GL is a redesign of RTISI-LA~\cite{RTISI-LA}. It is open sourced on Figure~\ref{fig:GL}. The algorithm receives as input a log magnitude spectrogram \textit{mag\_f} with size 1025, i.e., equal to the FFT size divided by two, plus one. Then, it inverts the natural logarithm by exponentiating the input magnitude frame (line 6 on Figure~\ref{fig:GL}). The magnitude spectrogram is converted to a complex-valued spectrogram by combining \textit{mag\_f} with zero phase. A sliding window queue \textit{mag\_w} is updated, by appending the current magnitude frame \textit{mag\_f} to the sequence of previously stored frames \textit{mag\_w} and then keeping the latest \textit{w\_size} frames. With this, \textit{mag\_w} always has a fixed number of \textit{w\_size} frames with the last dimension equal to 1025. A sliding window queue \textit{stft\_w} is updated with the current complex-valued spectrogram, as described in the previous step. We pre-compute phase of committed frames (in line 16) and use them as a phase constrain, so that phase of committed frames do not change during GL iterations below. A number of \textit{n\_iters} GL iterations are executed based on the current content of the sliding window queues (line 18). Namely, this consists of computing the inverse and forward STFT, estimating the uncommitted phase and recomputing \textit{stft\_w} by combining committed phase \textit{commit\_phase} and uncommitted phase \textit{uncommit\_phase} with the magnitude spectrogram \textit{mag\_w} (line 29 on Figure~\ref{fig:GL}). It allows to flow information between committed and uncommitted frames and use it for phase estimation of uncommitted frames in STFT domain. The output frame \textit{stft\_o} is extracted by reading the values of the STFT window queue \textit{stft\_w} at index \textit{ind}. Where \textit{ind} is an index of the current uncommitted frame in sliding window, so that all frames with indexes \textless \textit{ind} and indexes \textgreater \textit{ind} are committed and uncommitted (looking ahead) accordingly.

The algorithm described above is executed in streaming mode, whenever a new log magnitude spectrogram frame is available. Once \textit{stft\_o} is computed, a new frame of 200 samples of audio are synthesized running the streaming inverse STFT, according to the implementation in~\cite{KWS}. Streaming inverse STFT introduces an algorithmic delay equal to \textit{frame\_size} - \textit{frame\_step}. Finally, a de-emphasis filter is computed in the time domain to invert the pre-emphasis filter~\cite{HTK} (with \textit{coef}=0.97) and the final output is normalized by (1.0 + \textit{coef}).

We could not find open source version of original~\cite{RTISI-LA} and there was no subjective evaluation metrics reported in ~\cite{RTISI-LA}, so we can not compare our design with the original implementation of~\cite{RTISI-LA}.

In Section~\ref{experiments} we benchmark the streaming GL algorithm with window size \textit{w\_size}=4, \textit{n\_iters}=4 iterations and \textit{ind}=2. So that it uses 2 hops from the past and one hop from the future (one hop lookahead) for GL iterations, thus it produces an algorithm delay equal to 1 hop or 12.5ms. Additional delay is introduced by streaming inverse STFT (discussed above), so the total delay is equal to one frame. We label this approach as \textit{sGL1}. We compare it with non-streaming GL, labeled as \textit{nGL}, which takes the whole audio sequence and runs 70 iterations on it, thus it can not run in streaming mode.

\begin{figure}[t]
  \centering
\begin{lstlisting}[language=python]
# Initialize magnitude and STFT sliding window
mag_w = tf.zeros((1,w_size,1025),tf.float32)
stft_w = tf.zeros((1,w_size,1025),tf.complex64)
def streaming_griffin_lim(mag_f,mag_w, stft_w):
  # Inverse log
  mag_f = tf.exp(mag_f) - delta
  # Magnitude frame to complex frame
  stft_f = tf.complex(mag_f, 0.0) * tf.exp(
      tf.complex(0.0, tf.zeros_like(mag_f)))
  # Magnitude sliding window
  mag_w = tf.concat([mag_w, mag_f], 1)
  mag_w = mag_w[:, -w_size:, :]
  # STFT sliding window
  stft_w = tf.concat([stft_w, stft_f], 1)
  stft_w = stft_w[:, -w_size:, :]
  commit_phase = tf.math.angle(
      stft_w[:, 0:ind, :])
  for _ in range(n_iters):  # GL iterations
    audio_w = tf.signal.inverse_stft(
        stft_w, frame_size, frame_step, 
        fft_size, window_fn=None)  
    stft_w = tf.signal.stft(
        audio_w, frame_size, frame_step, 
        fft_size, window_fn=hann_window)  
    uncommit_phase = tf.math.angle(
        stft_w[:, ind:w_size, :])
    phase_w = tf.concat([commit_phase, 
        uncommit_phase], 1)
    stft_w = tf.complex(mag_w, 0.0) * tf.exp(
            tf.complex(0.0, phase_w))
  stft_o = stft_w[:,ind:ind+1,:] # Output frame
  return stft_o, mag_w, stft_w
\end{lstlisting}  
  \caption{Streaming GL}
  \label{fig:GL}
  \vspace{-3mm}
\end{figure}

\newcommand{\numblocks}{B}
\newcommand{\baseconvdepth}{C}
\newcommand{\ratio}{M}

\subsection{Streaming MelGAN} \label{subsectionNN}

We present streaming version of \textit{MelGAN}~\cite{kumar2019melgan}. Its models topology with parameters is the same with~\cite{kumar2019melgan} as shown in  Table~\ref{tab:model}. In streaming mode the input magnitude spectrogram with time dim 1 (this approach also supports processing multiple frames) and number of channels \textit{ch}=1025, is processed by a 1D convolution \textit{conv1D} with 512 channels and kernel size \textit{ks}=7. This is followed by four upscale1D blocks with channels~\textit{ch}, kernel size \textit{ks} and stride listed in Table \ref{tab:model}. Then, after the ELU activation function ~\cite{elu}, a final 1D convolution is applied with the number of channels \textit{ch}=1 and kernel size \textit{ks}=7. In Table~\ref{tab:model} we show how the time dimension and the number of channels are changing with every layer execution, so that in the last conv1D layer we get 200 audio samples with channel dimension 1, thus matching the hop size of the STFT (12.5ms at 16kHz sampling frequency). 

The structure of the \textit{upscale1D} block is shown in Table~\ref{tab:upscale1D}: after ELU activation, the signal is upscaled by a Conv1DTranspose layer with the following parameters: number of channels \textit{ch}, kernel size \textit{ks} and \textit{stride}, which are defined in Table~\ref{tab:model}. The upscaled signal is processed by a sequence of three residual blocks \textit{resBlk} with kernel size \textit{ks} and dilation shown in Table~\ref{tab:upscale1D}. 

The structure of the residual block \textit{resBlk} is shown in Table~\ref{tab:resBlk}: after ELU activation, the signal is processed by 1D convolution with kernel size \textit{ks}=3 and dilation shown in Table~\ref{tab:upscale1D} corresponding to \textit{resBlk}. Then, the output of the last 1x1 convolution in Table~\ref{tab:resBlk} is added to the input of \textit{resBlk} and returned as the final output of the \textit{resBlk}.

To guarantee real-time inference, all convolutions are \emph{causal} and running is streaming mode as in~\cite{li2021seanet}. The core streaming components are open sourced in~\cite{KWS}.
In order to provide lookahead, during training the target samples are shifted with respect to the input samples in the waveform domain, before computing the STFT. 

We train the neural vocoder (on 4 TPU with 1M training steps during 4.4 days) with the same mix of losses used in~\cite{soundstream2021} to achieve both signal reconstruction fidelity and perceptual quality, following the principles of the perception-distortion trade-off discussed in~\cite{blau2018perception}. The adversarial loss is used to promote perceptual quality and it is defined as a hinge loss over the logits of the discriminator, averaged over multiple discriminators and over time, operating both in the time domain and in the STFT domain. To promote fidelity of the decoded signal with respect to the input waveform as in HifiGAN~\cite{HIFI} we adopt two additional losses: i) a ``feature'' loss, computed in the feature space defined by the discriminator(s)~\cite{kumar2019melgan}; ii) a multi-scale spectral reconstruction loss~\cite{engel2020ddsp}.

\begin{table}[t]
  \begin{center}
    \caption{Model architecture.}
    \label{tab:model}
    \scalebox{0.9}{
    \begin{tabular}{l|c|c|c|c} 
      \textbf{operation} & \textbf{time dim} & \textbf{ch} & \textbf{ks} & \textbf{stride}\\
      \hline
      \textit{input}       & 1      & 1025  &     &    \\
      \textit{conv1D}      & 1      & 512   & 7   & 1  \\
      \textit{upscale1D}   & 5      & 256   & 10  & 5  \\      
      \textit{upscale1D}   & 25     & 128   & 10  & 5  \\      
      \textit{upscale1D}   & 100    & 64    & 8   & 4  \\      
      \textit{upscale1D}   & 200    & 32    & 4   & 2  \\      
      \textit{elu}         & 200    & 32    &     &    \\            
      \textit{conv1D}      & 200    & 1     & 7   & 1  \\            
    \end{tabular}
    }
    \vspace{-4mm}
  \end{center}
\end{table}

\begin{table}[t]
  \begin{center}
    \caption{upscale1D structure.}
    \label{tab:upscale1D}
    \scalebox{0.9}{
    \begin{tabular}{l|c|c|c|c} 
      \textbf{operation} & \textbf{ch} & \textbf{ks} & \textbf{stride} & \textbf{dilation}\\
      \hline
      \textit{elu}             & ch    &      &        &    \\            
      \textit{conv1Dtranspose} & ch    & ks   & stride & 1  \\      
      \textit{resBlk}          & ch    & 3    & 1      & 1  \\      
      \textit{resBlk}          & ch    & 3    & 1      & 3  \\      
      \textit{resBlk}          & ch    & 3    & 1      & 9  \\      
    \end{tabular}
    }
    \vspace{-4mm}
  \end{center}
\end{table}

\begin{table}[t]
  \begin{center}
    \caption{resBlk structure.}
    \label{tab:resBlk}
    \scalebox{0.9}{
    \begin{tabular}{l|c|c|c|c} 
      \textbf{operation} & \textbf{ch} & \textbf{ks} & \textbf{stride} & \textbf{dilation}\\
      \hline
      \textit{elu}           & ch    &      &      &           \\            
      \textit{conv1D}        & ch    & 3    & 1    & dilation  \\      
      \textit{conv1D}        & ch    & 1    & 1    & 1         \\      
    \end{tabular}
    }
    \vspace{-2mm}
  \end{center}
\end{table}

The training data consists of a mix of clean and noisy speech. We introduced noise in the training data, to make a model less sensitive to noisy speech (noise sensitivity will be observed in section \ref{experiments_subjective}. For clean speech, we use the LibriTTS dataset~\cite{zen2019libritts} with the following training splits: \emph{train-clean-100}, \emph{train-clean-360} and \emph{train-other-500}. For noisy speech, we synthesize samples by mixing speech from LibriTTS with noise from Freesound~\cite{fonseca2017freesound}. We apply peak normalization to randomly selected crops of 3 seconds and adjust the mixing gain of the noise component sampling uniformly in the interval $[-30 \, \text{dB}, 0 \, \text{dB}]$. 

In Section~\ref{experiments} we evaluate standard non streaming MelGAN~\cite{kumar2019melgan}, labeled as \textit{nMelGAN}. It has has the highest quality with high delay 187ms (it is hard to use for real time applications). We hypothesize that with only one hop delay we can get acceptable quality. So we introduce streaming  \textit{sMelGAN1} with lookahead 1 hop (12.5ms delay) and compare it with causal model with zero delay \textit{sMelGAN0}. These models have the same topology (defined in Table \ref{tab:model}) with 12M parameters.

\section{Experimental results} \label{experiments}

\subsection{Models benchmarks on mobile phone}  \label{experiments_mobile}
For real time applications it is important to have low delay (difference between time when signal is received and time when corresponding output is generated, excluding processing time: it is also called algorithmic delay), low streaming latency (time required to process one hop of audio in streaming mode). small model size (file size of TFLite module measured in MB) and memory footprint.
We benchmarked \textit{nGL}, \textit{sGL1}, \textit{sMelGAN1}, \textit{sMelGAN0} and \textit{nMelGAN} on single-threaded CPU of a Pixel4 CPU and reported results on on Table~\ref{tab:bench}. Models were executed with TFLite~\cite{TFL}. As expected Streaming GL has negligible TFLite model size (100KB), it is 250x times smaller than model size of the MelGAN model. We observe that \textit{sGL1} is 23\% faster. \textit{sGL1} uses only 7.6MB of CPU memory and it is 4.5x times less run-time memory than \textit{sMelGAN1} (34MB as shown on Table~\ref{tab:bench}). Non streaming models \textit{nGL} and \textit{nMelGAN} have the worst delay. As expected, causal model \textit{sMelGAN0} has zero delay but it has the worst quality as shown on Figure~\ref{fig:clean}. So we explore a trade-off between delay, model size and quality by introducing models \textit{sGL1} and \textit{sMelGAN1}.

\begin{table}[t]
  \begin{center}
      \caption{Delay, streaming latency, file and memory size}
      \label{tab:bench}
      \scalebox{0.9}{
      \begin{tabular}{p{1.4cm} | p{1.0cm}  | p{1.0cm}| p{1.0cm}| p{1.2cm}}
        \textbf{Models} & \textbf{Delay [ms]} & \textbf{Latency [ms]} & \textbf{FileSize [MB]} & \textbf{Memory [MB]}  \\
        \hline
          \textit{nGL}       & 2000   & N/A    & 0.1  \\        
          \textit{nMelGAN}   & 187   & N/A    & 25  \\
          \textit{sGL1}     & 12    & 5.2    & 0.1 & 7.6  \\
          \textit{sMelGAN1}  & 12   & 6.7    & 25 & 34  \\
          \textit{sMelGAN0}  & 0   & 6.7     & 25 & 34  \\          
      \end{tabular}
      }
  \end{center}
  \vspace{-4mm}
\end{table}

\subsection{Subjective quality evaluation} \label{experiments_subjective}

\begin{figure}[htbp!]
  \centering
  \includegraphics[width=\linewidth]{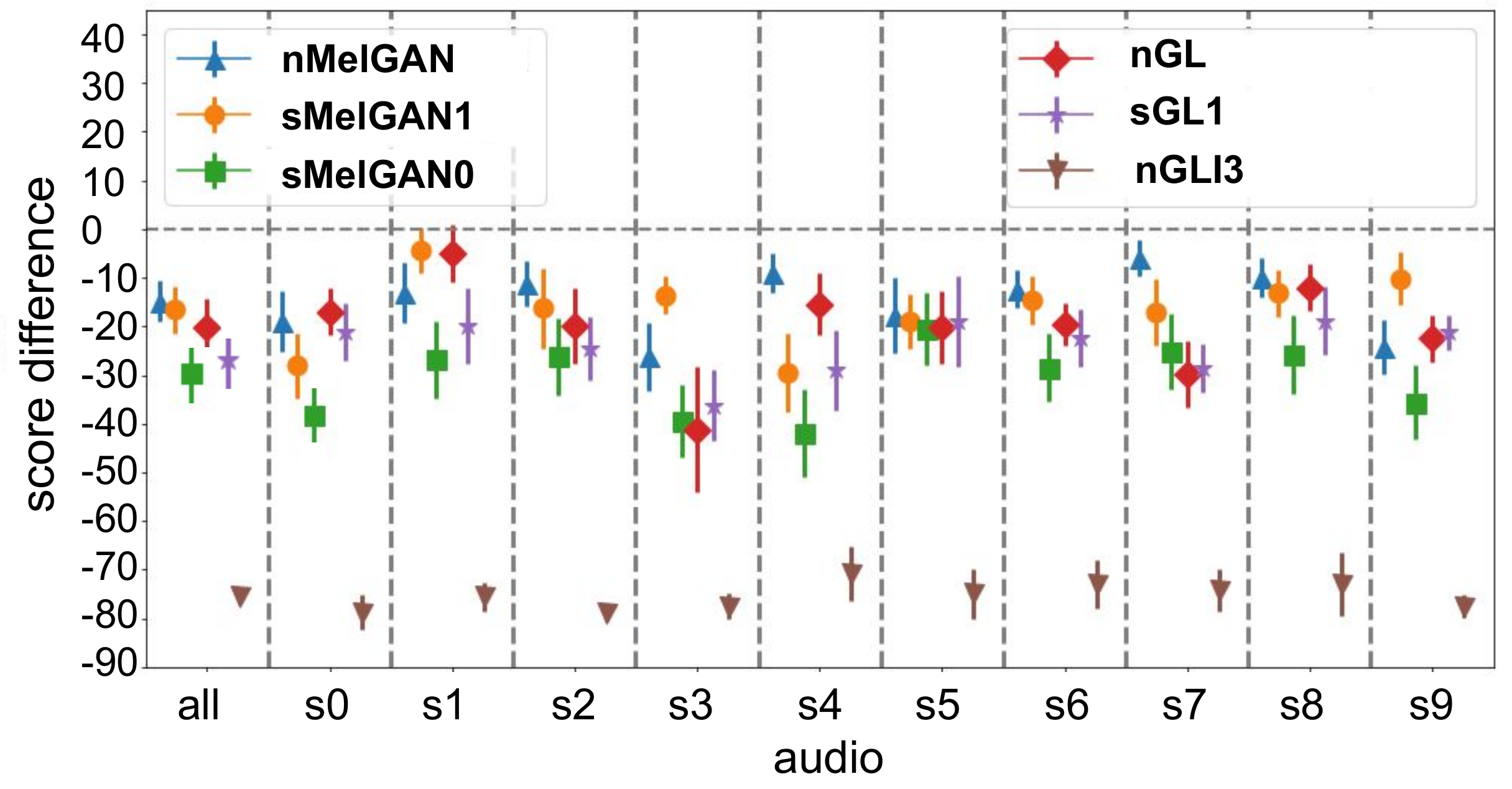}
  \caption{Average MUSHRA score differences on VCTK clean audio samples, between model outputs and the ground truth input audio, with 95\% confidence intervals.}
  \label{fig:clean}
  \vspace{-2mm} 
\end{figure}

We perform subjective evaluation using the MUSHRA methodology~\cite{mushra2015} with 10 test audio clips (2–5 seconds each) from the VCTK dataset~\cite{vctk} and show it on Figure~\ref{fig:clean}. Input audio is down-sampled to 16kHz, converted to magnitude spectrogram, and then inverted using the various approaches described in this paper. We compare the ground truth audio, together with \textit{nGL}, \textit{sGL1}, \textit{sMelGAN1}, \textit{nMelGAN}, \textit{sMelGAN0}. We run the evaluation on each of the 10 audio clips, which results in 10 groups of 7 audio clips (outputs of \textit{nGL}, \textit{sGL1}, \textit{sMelGAN1}, \textit{nMelGAN}, \textit{sMelGAN0} and ground-truth, where clips within a group have the same content but varying quality. The speakers in the 10 testing audio clips were not present in the training data of the neural vocoders. To calibrate our 10 raters, we first present them non-streaming GL with only 3 iterations \textit{nGLI3} (score = 20) with their corresponding ground truth audio (score = 100). We then ask the raters to assign scores between 0 and 100 to each of the 10x7 clips. 
We ranked the presented methods: 
\textit{nMelGAN} $\approx$ \textit{sMelGAN1} $>$ \textit{nGL} $>$ \textit{sGL1}  $\approx$  \textit{sMelGAN0}, using the Mann-Whitney U rank test (for p-value $\leq$ 0.01). We use sign $\approx$ to label pairs with p-value $>$ 0.01. The best model (in terms of quality-delay trade-off) on clean data is \textit{sMelGAN1}. With only one-hop lookahead, it significantly outperforms all GL approaches and the causal \textit{sMelGAN0}. Note that \textit{sGL1} has comparable quality with causal \textit{sMelGAN0}.

To assess the noise sensitivity of these vocoders, we repeated our subjective evaluation on the noisy version of the same 10 audio clips from VCTK~\cite{vctk}. In this case, the pairwise ranking comes out to be different from the clean-speech scenario. Again using the Mann-Whitney U rank test (for p-value $\leq$ 0.01), we observe \textit{nMelGAN} $\approx$ \textit{nGL} $\approx$ \textit{sMelGAN1} $\approx$ \textit{sGL1} $>$ \textit{sMelGAN0}. We believe that the quality reduction of the MelGAN is likely caused by the difference in noise distribution between training and testing time. GL approaches, on the other hand, are less sensitive to noise such that \textit{sGL1} and \textit{sMelGAN1} have similar perceptual quality on this test. In addition \textit{sGL1} outperforms causal \textit{sMelGAN0}, we explain it by 1 hop lookahead in \textit{sGL1}.  Hence \textit{sGL1} can be a better choice for faster inference (with less memory consumption), whereas \textit{sMelGAN1} would be a better option on clean audio and only one hop delay.  All data with the code to run demo models are open sourced at  link\footnote{\url{https://github.com/google-research/google-research/tree/master/specinvert}}.

\subsection{Objective quality evaluation: impact on WER of atypical speech conversion}
Our final goal is to run the vocoder in real production. For example, on production Parrotron application. Parrotron~\cite{Parrotron} is an end-to-end speech conversion model that is trained to convert atypical speech to fluent speech of a canonical synthesized voice. This model takes a log-mel spectrogram as an input and directly produces linear spectrogram. It requires a vocoder to synthesize a time-domain waveform in real time, thus a lightweight streaming vocoder is a crucial requirement for such an application. Parrotron normalizes atypical speech, so people with speech disabilities can communicate with others as well as speech enabled interfaces, such as Google Home, Siri or Amazon Alexa. In this case we need to see the WER metric explicitly. For example, state of the art Google ASR can have the WER on atypical speech more than 50\% (as shown in paper~\cite{Parrotron}), but if we pass atypical speech through the Parrotron model then WER can be reduced by 2-4x or more (so that Google Home or Amazon Alexa will able to understand normalized speech). Default production Parrotron model used non streaming GL (as a result it introduced additional several seconds delay impacting customer experience), in our paper we replaced it by streaming GL (reducing the delay down to 20ms) and evaluated the Parrotron model on atypical speech, and showed that streaming GL has no impact on WER and can be used in production. Here is a final demo~\cite{PARROTRON_DEMO_V} of production Parrotron with Streaming GL, presented in our paper (we can see that speech is easy to understand and normalized speech is generated as soon as the speaker finishes talking: due to streaming GL there is no delay).

In this section, we employ Google's state of the art ASR engine to automatically evaluate the vocoded Parrotron's output while testing our three vocoders. In Table~\ref{tab:parrotron} we report word error rate~(WER) of ASR engine by passing the converted speech from: a profoundly deaf speaker, a speaker with ALS, and a speaker with Muscular Dystrophy (MD). All those speakers are considered to have severe speech impairments.  Comparing WER across different vocoders is a good approximation to check if the linguistic content is preserved after vocoding the converted spectrogram. It is important to note that the absolute WER value is irrelevant here, as it evaluates how Parrotron normalizes atypical speech, not the vocoding phase. 

In Table~\ref{tab:parrotron} we observe that overall all vocoders preserve the linguistic content for all tested speakers. Interestingly, although the neural vocoder has not been tuned for auto generated spectrograms, it is still performing relatively well when compared to GL. Both GL approaches seem to perform similarly, but it is 1\% worse in one of the speakers when compared to the non-streaming GL. We can comfortably conclude that the use of either of streaming approaches are appropriate solution for speech conversion. Presented streaming vocoders can run in real time not only on mobile phone ARM CPU, but on cloud x86 CPU too, so streaming GL was launched in production for Parrotron cloud application, shown in demo~\cite{PARROTRON_DEMO_V}.

\begin{table}[t]
  \caption{Comparing WER from different vocoders after running Parrotron on atypical speech}
  \label{tab:parrotron}
  \scalebox{0.9}{
      \begin{tabular}{p{2.5cm} | p{1.5cm}  | p{1.5cm} | p{1cm}}
        \textbf{Model} & \textbf{Deaf} & \textbf{ALS} & \textbf{MD}\\
        \hline
          \textit{nGL}  & 21.2    & 22.5 &  10.4 \\    
          \textit{sGL1}       & 22.2    & 22.5 &  10.4 \\
          \textit{sMelGAN1}    & 20.6    & 23.0 &  10.9 \\      
      \end{tabular}
  }
  \vspace{-4mm}
\end{table}

\section{Conclusion}

We presented our design of streaming GL for inverting log magnitude spectrogram. It is 2.4x faster than real time (on a Pixel4 mobile phone ARM CPU), has only 0.1MB model size (vs MelGAN with 25MB), and similar quality with MelGAN on noisy audio data and real Parrotron application. This makes it attractive for wearable devices.

We also presented streaming MelGAN models and explored the impact of lookahead on the perceptual quality of generated audio. We showed that the model with even one-hop lookahead outperformed GL algorithms and causal MelGAN on clean audio clips. These models are also capable of real-time audio processing, achieving $\approx$2x real-time factor on Pixel4 CPU. However these MelGAN models need to be trained on speech data, and are more sensitive to noise vs GL approach. We showed that both streaming GL and streaming MelGAN are appropriate solutions for atypical speech conversion, e.g. Parrotron~\cite{Parrotron} application.

\bibliographystyle{IEEEtran}
{
\bibliography{mybib}
}

\end{document}